\documentclass{ieeeaccess}
\usepackage{cite}
\usepackage{amsmath,amssymb,amsfonts}
\usepackage{algorithmic}
\usepackage{graphicx}
\usepackage{textcomp}
\usepackage{lineno,hyperref}
\usepackage{graphicx}
\usepackage{float}
\usepackage{mathrsfs}
\usepackage{subfigure}
\usepackage{url}
\usepackage{multirow}
\usepackage{color}
\modulolinenumbers[5]
\usepackage{caption}
\usepackage{threeparttable}
\usepackage{indentfirst}
\usepackage[T1]{fontenc}
\usepackage{type1cm}
\usepackage{CJK}
\usepackage{bm}
\usepackage[inline]{trackchanges}
\usepackage{stfloats}
\usepackage{diagbox} 

\def\BibTeX{{\rm B\kern-.05em{\sc i\kern-.025em b}\kern-.08em
    T\kern-.1667em\lower.7ex\hbox{E}\kern-.125emX}}

\begin{document}
\history{Date of publication xxxx xx, xxxx, date of current March 2, 2020.}
\doi{10.1109/ACCESS.2020.DOI}

\title{W-net: Simultaneous segmentation of multi-anatomical retinal structures using a multi-task deep neural network}

\author{\uppercase{Hongwei Zhao\authorrefmark{1},Chengtao Peng\authorrefmark{3}}, 
\uppercase{Lei Liu$^\dagger$\authorrefmark{2}, Bin Li$^\dagger$\authorrefmark{3}}, Member, IEEE}

\address[1]{School of Information Science and Technology, University of Science and Technology of China, Hefei, Anhui, 230022, China}
\address[2]{Department of Precision Machinery and Instrumentation, University of Science and Technology of China, Hefei, Anhui, 230022, China}
\address[3]{CAS Key Laboratory of Technology in Geo-spatial Information Processing and Application System, University of Science and Technology of China, Hefei, Anhui, 230026, China}

\corresp{$\dagger$Corresponding author: Bin Li (e-mail: binli@ustc.edu.cn), Lei Liu (e-mail: liulei13@ustc.edu.cn)}

\markboth
{Zhao \headeretal:W-net: Simultaneous segmentation of multi-anatomical retinal structures using a multi-task deep neural network}
{Zhao \headeretal:W-net: Simultaneous segmentation of multi-anatomical retinal structures using a multi-task deep neural network}

\tfootnote{
This work is partially supported by the National Natural Science Foundation of China (Key Program) under grand No.U19B2044. It is supported by the GPU Computing Cluster of the data center, School of Information Science and Technology, University of Science and Technology of China. }

\begin{abstract}
Segmentation of multiple anatomical structures is of great importance in medical image analysis. In this study, we proposed a $\mathcal{W}$-net to simultaneously segment both the optic disc (OD) and the exudates in retinal images based on the multi-task learning (MTL) scheme. We introduced a class-balanced loss and a multi-task weighted loss to alleviate the imbalanced problem and to improve the robustness and generalization property of the $\mathcal{W}$-net. We demonstrated the effectiveness of our approach by applying five-fold cross-validation experiments on two public datasets e\_ophtha\_EX and DiaRetDb1. We achieved F1-score of 94.76\% and 95.73\% for OD segmentation, and 92.80\% and 94.14\% for exudates segmentation. To further prove the generalization property of the proposed method, we applied the trained model on the DRIONS-DB dataset for OD segmentation and on the MESSIDOR dataset for exudate segmentation. Our results demonstrated that by choosing the optimal weights of each task, the MTL based $\mathcal{W}$-net outperformed separate models trained individually on each task. Code and pre-trained models will be available at: \url{https://github.com/FundusResearch/MTL_for_OD_and_exudates.git}.
\end{abstract}
\begin{keywords}
Multi-task Learning, Unet, Optic Disc, Exudates
\end{keywords}

\titlepgskip=-15pt

\maketitle

 \section{INTRODUCTION}
 \label{sec:intro}  
 \noindent It has been long that philosophers defined the eye as the window to the human soul. Indeed, more and more scientific and clinical evidences show that retina is indeed an extension of the central nervous system (CNS). During embryonic development, the retina and optic nerve extend from the diencephalon, and are thus considered part of the CNS~\cite{London:2013kt}. The morphological variation in the retinal anatomical structures is of great diagnostic value, not only for the retinal pathology such as diabetic retinopathy (DR)~\cite{sopharak2008automatic}, glaucoma~\cite{joshi2011optic}, and age-related macular degeneration (AMD)~\cite{lim2012age}, but also for the systemic heart- and brain-related diseases, such as hypertension, Parkinson's, and Alzheimer's diseases (AD)~\cite{McGrory:2017iq}. Since imaging of the retina with fundus photographs is non-invasive and inexpensive, many studies have been conducted to develop ocular biomarkers for AD or Parkinson's disease in the hope of potential wide implementation. Quantitative analysis of the retinal anatomic structures, such as the optic disc (OD), optic cup (OC), blood vessels, and other pathological features like exudates and hemorrhage, is the first and essential step in the development of automated diagnostic or screening system.
 
 OD segmentation is an essential step in detecting glaucoma, a chronic eye disease in which the optic nerve is gradually damaged. It is the second leading cause of blindness, with the global prevalence of 3.54\% for population aged 40-80 years old~\cite{tham2014global}. If undiagnosed, glaucoma causes irreversible damage to the optic nerve leading to blindness. Therefore diagnosing glaucoma at early stages is extremely important for appropriate management and treatment of the disease~\cite{hitchings2017european}. The cup to disc ratio (CDR) is widely accepted and commonly used by clinicians to screen for glaucoma. The accurate segmentation of OD is critical to determine CDR. 
 There are studies on the automatic segmentation of OD in retinal images, which can be mainly grouped into four categories: morphological-based approaches~\cite{aquino2010detecting}, pixel classification methods~\cite{cheng2013superpixel}, deformable model methods~\cite{giachetti2014accurate}, and deep learning methods~\cite{fu2018joint,edupuganti2018automatic}. In the group of morphological-based methods, Aquino \textit{et al.}~\cite{aquino2010detecting} presented a template-based methodology for segmenting the OD by using morphological and edge detection techniques followed by the circular Hough transformation. Another group of methods was proposed to transfer the boundary detection problem into a pixel classification task. Cheng \textit{et al.}~\cite{cheng2013superpixel} used histograms and the center surround statistics to classify each pixel as disc or non-disc. Based on deformable model, Giachetti \textit{et al.}~\cite{giachetti2014accurate} used ellipse fitting combined with a radial symmetry detector and a vessel density map to segment OD.  
 
 Deep convolutional neural network (DCNN) has been applied recently in retinal image processing. Huazhu Fu \textit{et al.}~\cite{fu2018joint} proposed a deep learning architecture (M-net), and Venkata Gopal Edupuganti \textit{et al.}~\cite{edupuganti2018automatic} applied a fully convolutional neural network (FCN) to solve OD segmentation problem.
 
\begin{figure}[htpb]
 \centering
 \includegraphics[width=.85\linewidth]{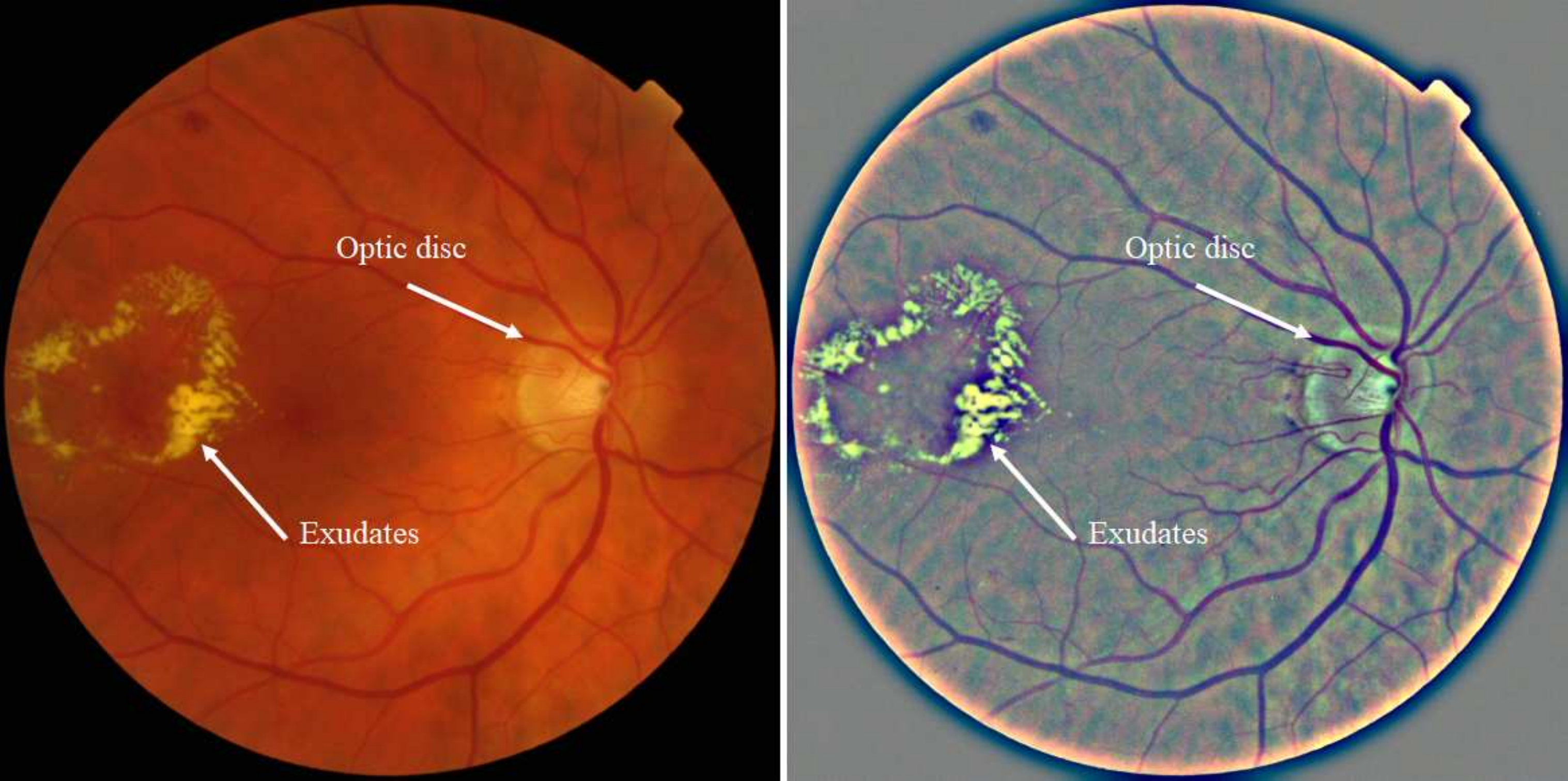}
 \caption{(a) A retina image with labeled exudates and the optic disc. (b) The preprocessed image of (a) using Eq.(\ref{Eq:eq_gussain1}).}
 \label{fig:retinal}
 \end{figure}

 Exudates are the early signs of diabetic retinopathy (DR), which caused 2.6\% of blindness worldwide in 2010~\cite{bourne2013causes}. Exudates are lipid and lipoprotein deposits that appear near leaking capillaries within the retina. They develop at the early stage of DR and may appear as yellow areas with variable sizes from a few pixels to as large as the optic disc (Fig.~\ref{fig:retinal}(a)). After 20 years of diabetes, nearly all patients with Type I diabetes and >60\% of patients with Type II diabetes have some degree of retinopathy~\cite{american1966diabetes}. Though the exudates and OD may look very different, they do share some similarity, for example, both of them appear as bright yellow areas in color fundus photographs. Thresholding methods, based on either global or local image gray-levels~\cite{phillips1993automated,wisaeng2015automatic}, or clustering-based image thresholding, such as Otsu thresholding~\cite{wisaeng2015automatic}, had been applied for exudate segmentation. J. Kaur \textit{\textit{et al.}}~\cite{kaur2018generalized} proposed a dynamic decision thresholding method to reliably segment exudates irrespective of the associated heterogeneity, brightness and faint edges of the image. Region growing methods~\cite{ege2000screening} were demonstrated for automatic segmentation of exudates, particularly in combination with the artificial neural network~\cite{usher2004automated}. In the category of machine learning based methods, Benalc{\'a}zar \textit{et al.}~\cite{benalcazar2013automatic} used logistic regression to classify each exudate candidate detected by the ensemble of aperture filters.  Giancardo \textit{et al.}~\cite{giancardo2012exudate} segmented exudates using features extracted manually in combination with support vector machine (SVM). 
 
 DCNNs have been proven to outperform conventional image analysis methods in many aspects, especially because they do not require explicit feature extraction. Recently, deep learning methods were applied to detect exudates and OD in color fundus photographs and showed promising results~\cite{feng2017deep,srivastava2015using}. However, these methods delineated one kind of anatomy at a time.
 
 In contrast, multi-task learning~\cite{caruana1997multitask} seeks to simultaneously learn a set of task-specific classification or regression models. The rationale behind MTL is that by utilizing the correlation information among the related tasks, a joint learning method is much more efficient than learning each task separately.  
 MTL has been successfully employed in image classification~\cite{yuan2012visual}, visual tracking~\cite{zhang2013robust}, multi-view action recognition~\cite{yan2014multitask} and egocentric daily activity recognition~\cite{yan2015egocentric}. However, to the best of our knowledge, there was no study of using MTL in the retinal anatomy segmentation. 
 
  In the study, based on the multi-task deep learning scheme, we constructed a $\mathcal{W}$-net to simultaneously segment both OD and exudates. By utilizing the correlation between the tasks, the network showed better performance than individual one-task only networks. This method was ready to apply to other multi-task scenarios, especially in the field of medical image analysis.

 \section{Materials and Methods}
 \noindent Based on the MTL scheme, we proposed a $\mathcal{W}$-net to simultaneously segment OD and exudates. We trained our model using two datasets: e\_ophtha\_EX~\cite{zhang2014exudate} and DiaRetDb1~\cite{kauppi2007diaretdb1}, but tested on four datasets: e\_ophtha\_EX~\cite{zhang2014exudate}, DiaRetDb1~\cite{kauppi2007diaretdb1}, DRIONS-DB~\cite{carmona2008identification} and MESSIDOR~\cite{decenciere2014feedback}. 
  
  \subsection{Datasets}
 \noindent The e\_ophtha\_EX dataset consists of 82 fundus images. There are four different image sizes of 1440$\times$960 or 1504$\times$1000 or 2544$\times$1696 or 2048$\times$1360 pixels. All the images were acquired with a 45-degree field of view. The dataset contains 47 images with exudates and 35 images with no lesion. This dataset does not provide the ground truth for the optic discs. With the help of an ophthalmologist, we manually annotated the OD pixels for all the 82 images. 
 
 The DiaRetDb1~\cite{kauppi2007diaretdb1} dataset contains 89 color fundus images, which were captured using the same 50 degree field-of-view digital fundus camera with varying imaging settings. There are 26 images containing exudates with annotated ground truth~\cite{kalviainen2007diaretdb1}. We manually annotated the OD pixels of all the 89 images.
 
 The DRIONS-DB dataset~\cite{carmona2008identification} consists of 110 images of 600$\times$400 pixels. All the ground truths of the OD pixels are provided. This public dataset was widely used to evaluate the performance of OD segmentation algorithms. 
 
 The MESSIDOR\cite{decenciere2014feedback} dataset consist of 1200 TIFF images with three different image sizes, 1440$\times$960, 2240$\times$1488 and 2304$\times$1536 pixels. All the images were acquired with a color video CDD camera on a TopCon TRC NW6 with a 45-degree field of view. There are 226 images consisting of exudates and 974 images without exudates. We used 226 images with exudates and groundtruth images from~\cite{zheng2018detection}. 
 
 \subsection{Image preprocessing}
 \noindent All the images were rescaled to 576$\times$576 pixels. If the original images were smaller (for example images in the DRIONS-DB dataset), we applied zero-padding. The preprocessed image ${I}_{p}$ was evaluated using Eq.(\ref{Eq:eq_gussain1})~\cite{chudzik2018microaneurysm}
 
 \begin{equation}
 \mathcal{I}_{p} = \mathcal \alpha \cdot {I} + \mathcal \beta \cdot {I}_{Gauss} + \mathcal \gamma
 \label{Eq:eq_gussain1}
 \end{equation}
 
 \noindent where $\alpha$, $\beta$, and $\gamma$ were set to 4, -4, and 128, respectively. ${I}_{Gauss}$, the Gaussian blurred image was obtained with a 9$\times$9 Gaussian kernel. 
 
 \subsection{$\mathcal{S}$-net and $\mathcal{W}$-net structures}
\begin{figure}[htp!]
 \centering
 \includegraphics[width=.85\linewidth]{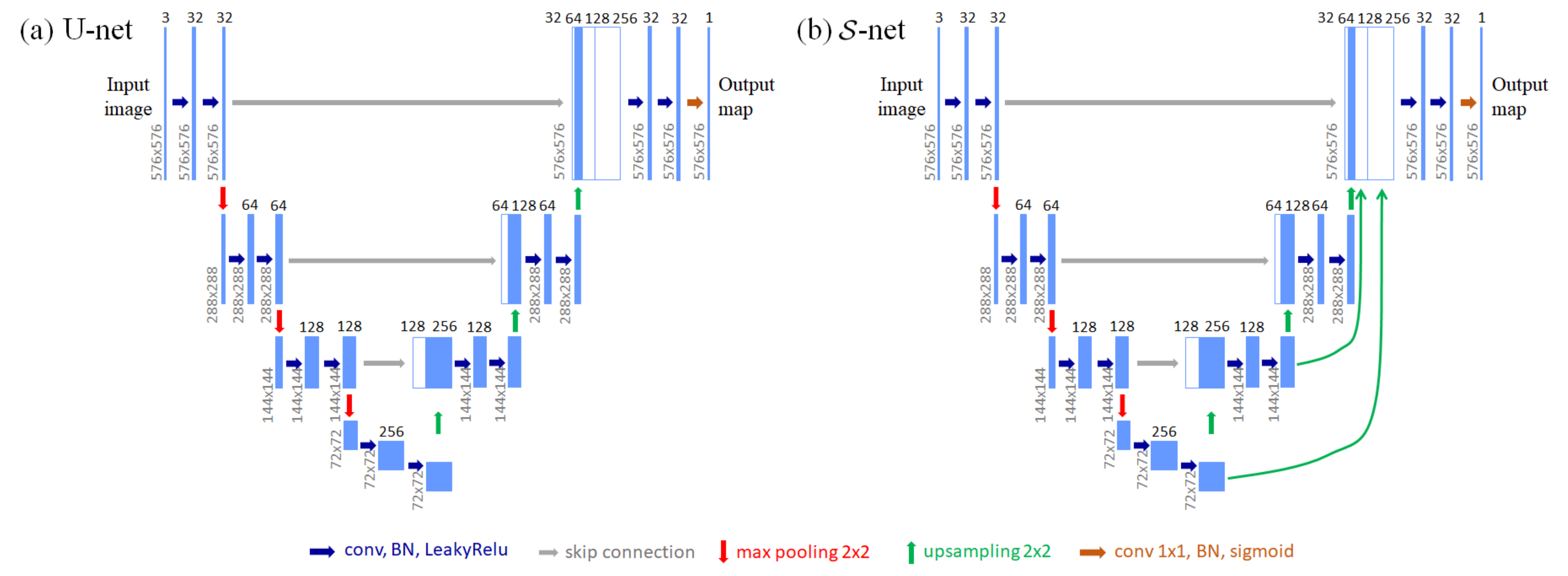}
 \caption{(a) An U-net structure. (b) The proposed single-task network structure ($\mathcal{S}$-net). The network was improved based on an U-net. In the decoder path of $\mathcal{S}$-net, the output of each upsampling layer was concatenated with the output of the last upsampling layer.}
 \label{fig:single}
 \end{figure}

 \noindent We constructed a single task network ($\mathcal{S}$-net),  which was based on a U-net~\cite{ronneberger2015u} (Fig.~\ref{fig:single}(a)). As shown in Fig.~\ref{fig:single}(b), the $\mathcal{S}$-net had an encoder and a decoder. The encoder path was responsible for the feature extraction, same as the U-net, while the decoder path was different with U-net in that every output of the upsampling layer was concatenated with the output of the last upsampling layer.  
 
\begin{figure}[htp!]
 \centering
 \includegraphics[width=.85\linewidth]{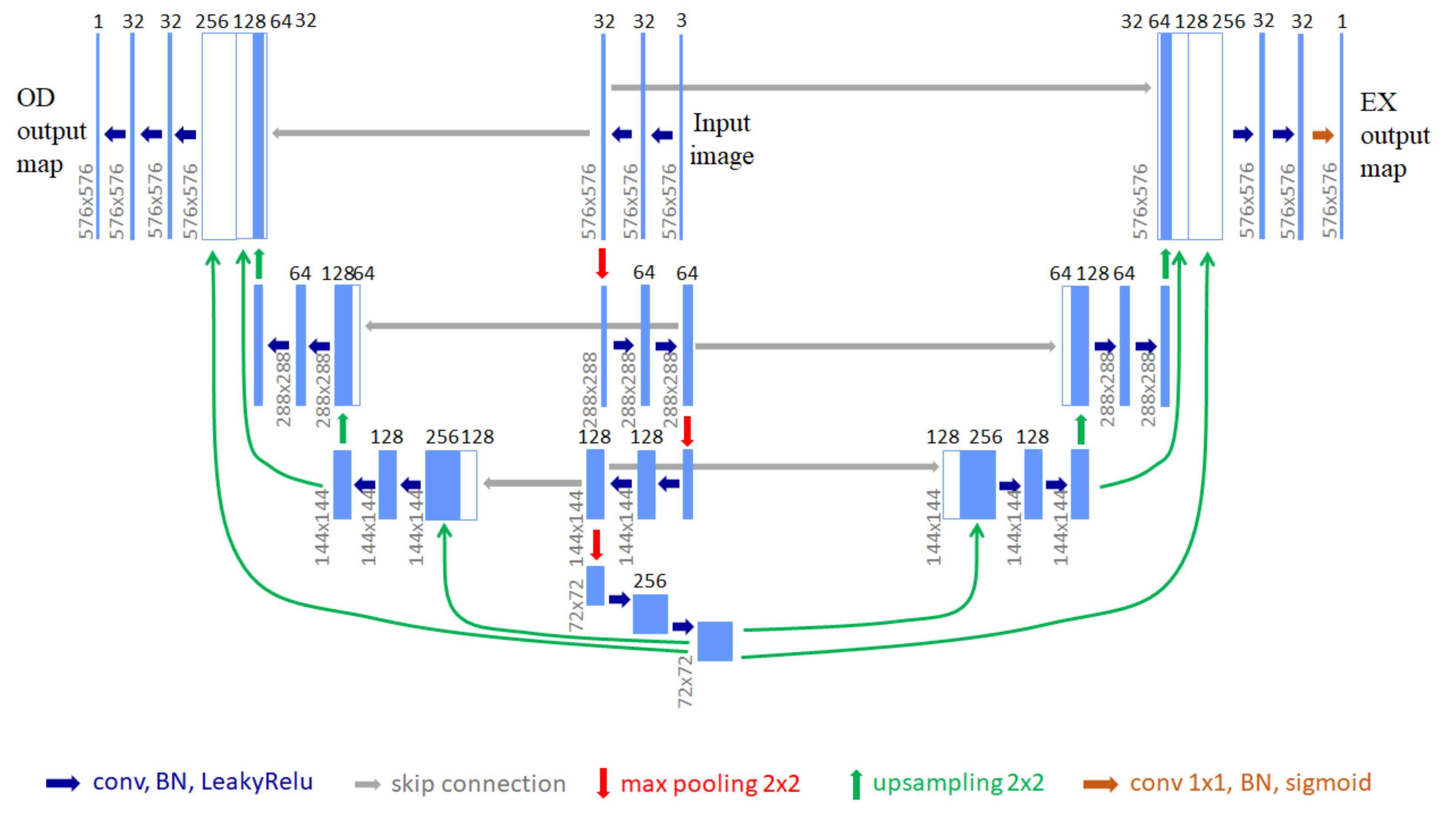}
 \caption{The structure of the proposed $\mathcal{W}$-net. The encoder (middle) was shared by the two decoders, which were used to segment the OD and exudates, respectively. Like in the U-net, the corresponding encoder layers and the decoder layers were skip connected. In addition, the same as the $\mathcal{S}$-net, the output of each upsampling layer was concatenated to the last upsampling layer in the decoder.} 
 \label{fig:structure}
 \end{figure}

 The proposed multi-task $\mathcal{W}$-net was based on the $\mathcal{S}$-net by adding a second decoder path for the second task, as shown in Fig.~\ref{fig:structure}. The two decoders shared the same encoder and generated the OD and exudates segmentations, respectively. The details of the configuration of the network were listed in Table~\ref{table:w-net structure}. In the encoder path, two convolutional layers, both with a 3$\times$3 kernel, were followed by normalization and dropout with a rate of 0.2. A leaky rectified linear unit (LeakyReLU)  was implemented as the activation function. A 2$\times$2 max pooling with a stride of two at each dimension was applied. In the decoder path, similar to the encoder path, two 3$\times$3 convolutional layers were followed by a LeakyReLU activation. An upsampling layer with a kernel size of 2$\times$2 and a stride of two in each dimension was applied and the output was fused with the corresponding layers in the encoder path. In the last layer, a 1$\times$1 convolution reduced the number of output channels to 2. The input of the network was 576$\times$576 and the output image size was the same as the input. In the decoder path of $\mathcal{S}$-net, the output of each upsampling layer was concatenated with the output of the last upsampling layer.
 
 \begin{table*}[htp!]
 \centering
 \begin{threeparttable}
 \small
 \setlength\tabcolsep{2pt}
 \centering 	 	
 \caption{The detailed configurations of the $\mathcal{W}$-net}
 \begin{tabular}{|l|c|l|c|l|c|}
 \hline
 \rule[-1ex]{0pt}{3.5ex}  Encoder path & Output size & Decoder path1 & Output size & Decoder path2 & Output size\\
 \hline
 \rule[-1ex]{0pt}{3.5ex} Conv1 32 3$\times$3 & 576$\times$576 & & & &\\
 \hline
 \rule[-1ex]{0pt}{3.5ex} Conv2 32 3$\times$3 & 576$\times$576 & A\_Conv7 2 1$\times$1 & 576$\times$576& B\_Conv7 2 1$\times$1 & 576$\times$576\\
 \hline
 \rule[-1ex]{0pt}{3.5ex} Max pooling 2$\times$2  & 288$\times$288 & A\_Conv6 32 3$\times$3 & 576$\times$576 & B\_Conv6 32 3$\times$3 & 576$\times$576\\
 \hline
 \rule[-1ex]{0pt}{3.5ex} Conv3 64 3$\times$3 & 288$\times$288 & A\_Conv5 32 3$\times$3 & 576$\times$576 & B\_Conv5 32 3$\times$3 & 576$\times$576\\
 \hline
 \rule[-1ex]{0pt}{3.5ex} Conv4 64 3$\times$3 & 288$\times$288 & Upsampling 2$\times$2 & 576$\times$576 & Upsampling 2$\times$2 & 576$\times$576\\
 \hline
 \rule[-1ex]{0pt}{3.5ex} Max pooling 2$\times$2  & 144$\times$144 & A\_Conv4 64 3$\times$3 & 288$\times$288  & B\_Conv4 64 3$\times$3 & 288$\times$288\\
 \hline
 \rule[-1ex]{0pt}{3.5ex} Conv5 128 3$\times$3 & 144$\times$144 & A\_Conv3 64 3$\times$3  & 288$\times$288& B\_Conv3 64 3$\times$3 & 288$\times$288  \\
 \hline
 \rule[-1ex]{0pt}{3.5ex} Conv6 128 3$\times$3 & 144$\times$144 & Upsampling 2$\times$2 & 288$\times$288 & Upsampling 2$\times$2 & 288$\times$288  \\
 \hline
 \rule[-1ex]{0pt}{3.5ex} Max pooling 2$\times$2  & 72$\times$72 & A\_Conv2 128 3$\times$3 & 144$\times$144  & B\_Conv2 128 3$\times$3 & 144$\times$144\\
 \hline
 \rule[-1ex]{0pt}{3.5ex} Conv7 256 3$\times$3 & 72$\times$72 & A\_Conv1 128 3$\times$3 & 144$\times$144  & B\_Conv1 128 3$\times$3 & 144$\times$144\\
 \hline
 \rule[-1ex]{0pt}{3.5ex} Conv8 256 3$\times$3 & 72$\times$72 & Upsampling 2$\times$2 & 144$\times$144 & Upsampling 2$\times$2 & 144$\times$144  \\
 \hline
 \end{tabular}
 \label{table:w-net structure}
 \end{threeparttable}
 \end{table*}
 
 \subsection{Loss functions}
 \noindent We defined the multi-task total loss $\mathcal{L}_{total}(x,\theta)$, the OD task loss $ \mathcal {L}_{OD}(x;\theta)$, and the exudate task loss $\mathcal {L}_{EX}(x;\theta)$ . 
 
 For easy reading, we applied the same symbols as used in the study by Maninis \textit{et al.}~\cite{maninis2016deep}. Briefly, the training dataset was denoted as $S = {(X_{m},Y_{m},Z_{m}), m=1,...,M}$, where $M$ was the total number of images with $X_{m}$ being the input image. $Y_{m} = \{y_{i}^{(m)}| i=1,...,N,|X_{m}, y_{i}^{(m)} \in\{0,1\}\}$ and $Z_{m} = \{z_{i}^{(m)}| i=1,...,N,|X_{m}, z_{i}^{(m)} \in\{0,1\}\}$ were the predicted pixel-wise labels of optic disc and exudates, respectively. The loss functions were then defined as:
 
 \begin{equation}
 \mathcal{L}_{total}(x,\theta) =  \mathcal \omega\cdot{L}_{EX}(x;\theta) + \mathcal (1-\omega)\cdot \mathcal {L}_{OD}(x;\theta)
 \label{Eq:total_loss1}
 \end{equation}
 
 \begin{equation}
 \begin{split}
 \mathcal{L}_{EX} = -\frac{1}{m}\sum_{m=1}^{M} [ -\lambda_2\sum_{i\in Z_{+}}\log P(z_{i}^{(m)}=1|X^{(m)};\theta) - \\ (1-\lambda_2)\sum_{i\in Z_{-}}\log P(z_{i}^{(m)}=0|X^{(m)};\theta) ]
 \label{Eq:eq_EX_loss}
 \end{split}
 \end{equation}
 
 \begin{equation}
 \begin{split}
 \mathcal{L}_{OD} = -\frac{1}{m}\sum_{m=1}^{M} [ -\lambda_1\sum_{i\in Y_{+}}\log P(y_{i}^{(m)}=1|X^{(m)};\theta) - \\ (1-\lambda_1)\sum_{i\in Y_{-}}\log P(y_{i}^{(m)}=0|X^{(m)};\theta) ]
 \label{Eq:eq_OD_loss}
 \end{split}
 \end{equation}
 
 \noindent where $\theta$ denoted the parameters of our $\mathcal{W}$-net. $Y_{+}$ and $Y_{-}$ represented the positive and negative label sets of the optic disc, respectively. $Z_{+}$ and $Z_{-}$ represented the exudates and non-exudates ground truth label sets, respectively. The probability $P(y_{i}^{(m)}=1|X^{(m)};\theta) \in [0,1]$ was computed using sigmoid function on the activation value at each pixel. We used $\lambda_1=0.7$ and $\lambda_2=0.9$ to alleviate the imbalance of the substantially greater number of background compared to the foreground pixels. It is important to choose the correct $\lambda_1$ and $\lambda_2$ in Eq.( \ref{Eq:eq_EX_loss}, \ref{Eq:eq_OD_loss}). $\mathcal{\omega}$ was the loss equilibrium parameter, which balanced the two different tasks. We applied the method proposed by Xie \textit{et al.}~\cite{xie2015holistically}, which used the ratio of the number of the background pixels and the number of pixels of the whole image.
 
 \subsection{Training and testing of the networks}
 \noindent We adapted five-fold cross-validation to evaluate the generalization property of the networks. Each dataset was splitted into 5 subsets of nearly equal size. Five independent models of each of the U-net, $\mathcal{S}$-net, and $\mathcal{W}$-net were trained. The models were trained on 4 splits (80\%) and tested on the remaining split (20\%) to assess the ability to generalize to the previously unseen images.
 
 During training, we did data augmentation by rotating the image with a random angle, flipping image randomly or adding random Uniform noise(U(0.8, 1.2)) to the image. In addition, the images were normalized by subtracting the averages of the pixels from each image and divided by the standard deviation.  We initialized the weights in each layer from a zero-mean Gaussian distribution with a standard deviation of 0.01($\mathcal{N}(0, 0.01)$). We trained the networks for 1000 epochs with a batch size of 2, and with an Adam optimizer with a learning rate of 0.0005 and $\mathcal{\beta}$ of 0.5.

 \section{Results}
 
 \subsection{Evaluation metrics}
 \noindent We evaluated the performance of the network both on the exudate and the OD segmentation. We applied the lesion level measurement for the exudates~\cite{wolf2006object, zhang2014exudate}, and the overlapping score metric for the OD segmentation~\cite{aquino2010detecting}. For the lesion level evaluation, we compared the predicted candidates with the ground truth provided in the datasets. We applied the approach based on set operations~\cite{wolf2006object, zhang2014exudate}. Briefly, if the two sets had enough overlap as controlled by an overlapping factor $\sigma$, which was set to 0.2 in our study, the same as  the work~\cite{zhang2014exudate}, the candidates were considered correctly classified. We computed the sensitivity, precision, and the F1-score according to the equations defined in Table~\ref{table:evaluation_metric}.
 
 \begin{table}[htp!]
 \centering
 \begin{threeparttable}
 \centering
 \caption{Definitions of the evaluation metrics}
 \begin{tabular}{ll}
 \hline
 \rule[-1ex]{0pt}{3.5ex} Performance Measure & Mathematical Formula \\
 \hline
 \rule[-1ex]{0pt}{3.5ex} Accuracy & (TP+TN)/(TP+TN+FN+FP) \\
 \rule[-1ex]{0pt}{3.5ex} Sensitivity & TP/(FN+TP) \\
 \rule[-1ex]{0pt}{3.5ex} Specificity& TN/(FP+TN)  \\
 \rule[-1ex]{0pt}{3.5ex} Precision & TP/(FP+TP)  \\
 \rule[-1ex]{0pt}{3.5ex} F1-Score & 2*TP/(2*TP+FP+FN)\\
 \hline
 \end{tabular}
 \label{table:evaluation_metric}
 \begin{tablenotes}
 \item[1] TP stands for true positive. FP: false positive; TN: true negtive; FN: false negative.
 \end{tablenotes}
 \end{threeparttable}
 \end{table}

 In order to compare our OD segmentation results with previous studies, we applied the overlapping score. Overlapping score ($\eta$) was a common evaluation metric for OD segmentation assessment~\cite{aquino2010detecting}. $\eta$ was defined as:
 
 \begin{equation}
 \eta = \frac{Area(P \cap G)}{Area(P \cup G)}
 \label{Eq:overlap}
 \end{equation}
 where $P$ and $G$ denoted the predicted and manually annotated OD regions, respectively. 
 
 \subsection{Comparisons of the $\mathcal{S}$-net and the U-net}
 \noindent To compare the performance of the $\mathcal{S}$-net and the U-net, we conducted the OD segmentation and the exudate segmentation on the e\_ophtha\_EX and DiaRetDb1 datasets. As shown in Table~\ref{tab:OD seg}, for the OD segmentation, on the e\_ophtha\_EX dataset, the U-net was slightly better than the $\mathcal{S}$-net, while on the DiaRetDb1 dataset, the $\mathcal{S}$-net was slightly better. However, for the exudate segmentation, as listed in Table~\ref{tab:exudates seg}, the F1-score achieved by the $\mathcal{S}$-net was higher than that by the U-net ($p=0.08$). Therefore, we chose the $\mathcal{S}$-net as the building block for the $\mathcal{W}$-net. 
 
 \begin{table*}[htp!]
 \centering
 \caption{Results of the OD segmentation using the U-net and the $\mathcal{S}$-net on the two datasets}
 \begin{tabular}{c|c|cccc}
 \multicolumn{2}{c|}{Experiments} & F1-score & Sensitivity & Precision & $\eta$\\
 \hline
 \rule[-1ex]{0pt}{3.5ex} \multirow{4}*{e\_ophtha\_EX} &  \multirow{2}*{U-net} & \bm{$93.84\% $} & \bm{$92.81\%$} & $\bm{94.91\%}$ & \bm{$88.45\% $} \\
 \rule[-1ex]{0pt}{3.5ex} ~ &  ~ & \bm{$ \pm 1.71\% $} & \bm{$ \pm 1.86\%$ }& $\bm{ \pm 1.64\%}$ & \bm{$ \pm 2.97\%$} \\
 \cline{2-6}
 \rule[-1ex]{0pt}{3.5ex} ~  &  \multirow{2}*{$\mathcal{S}$-net} & $93.46\% $ & $92.31\% $ & $94.69\% $& $87.84\%$ 
 \\
 \rule[-1ex]{0pt}{3.5ex} ~ &  ~ &  $\pm 2.79\% $ & $\pm 4.20\%$ & $ \pm 1.31\%$& $ \pm 4.75\%$ 
 \\
 \hline
 \rule[-1ex]{0pt}{3.5ex} \multirow{4}*{DiaRetDb1} & \multirow{2}*{U-net} & $95.33\%  $ & $94.65\% $ & $96.03\% $ & $91.08\% $\\
 \rule[-1ex]{0pt}{3.5ex} ~ & ~ & $\pm 0.78\% $ & $ \pm 1.21\%$ & $\pm 0.79\%$ & $\pm 1.44\%$\\
 \cline{2-6}
 \rule[-1ex]{0pt}{3.5ex}  ~  &  \multirow{2}*{$\mathcal{S}$-net} & \bm{$95.42\% $} & \bm{$94.76\% $} & \bm{$96.10\%$} & \bm{$91.25\% $}\\
 \rule[-1ex]{0pt}{3.5ex}  ~  &  ~ & \bm{$ \pm 0.81\% $} & \bm{$ \pm 1.66\%$} & \bm{$\pm 0.54\%$} & \bm{$ \pm 1.48\%$}\\
 \hline
 \end{tabular}
 \label{tab:OD seg}
 \end{table*}
 
 \begin{table*}[htp!]
 \centering
 \caption{Results of the exudate segmentation using the U-net and the $\mathcal{S}$-net on the two datasets}
 \begin{tabular}{c|c|ccc}
 \hline
 \multicolumn{2}{c|}{Experiments} & F1-score & Sensitivity & Precision \\
 \hline
 \rule[-1ex]{0pt}{3.5ex} \multirow{2}*{e\_ophtha\_EX} &U-net  & $91.14\% \pm 0.94\% $ & \bm{$92.87\% \pm 2.81\%$} & $89.57\% \pm 1.65\%$ \\
 \cline{2-5}
 \rule[-1ex]{0pt}{3.5ex} ~ & $\mathcal{S}$-net & \bm{$91.57\% \pm 1.01\%$}& $92.74\% \pm 1.26\%$ & \bm{$90.47\% \pm 1.95\%$} \\
 \hline
 \rule[-1ex]{0pt}{3.5ex} \multirow{2}*{DiaRetDb1} & U-net & $90.68\% \pm 2.63\% $ & $89.21\% \pm 3.52\%$ & $92.28\% \pm 2.76\%$ \\
 \cline{2-5}
 \rule[-1ex]{0pt}{3.5ex}          ~       & $\mathcal{S}$-net & \bm{ $91.25\% \pm 2.59\%$} & \bm{$89.57\% \pm 3.31\%$ }& \bm{$93.06\% \pm 2.86\%$ }\\
 \hline
 \end{tabular}
 \label{tab:exudates seg}
 \end{table*}
 
 \subsection{The $\mathcal{W}$-net improved both the OD and exudate segmentation performance}
 \begin{table*}[htp!]
 \centering
 \caption{Results of the OD segmentation using the $\mathcal{S}$-net and $\mathcal{W}$-net on the two datasets}
 \begin{tabular}{c|c|cccc}
 \hline
 \multicolumn{2}{c|}{Experiments} & F1-score & Sensitivity & Precision & $\eta$\\
 \hline
 \rule[-1ex]{0pt}{3.5ex} \multirow{4}*{e\_ophtha\_EX} & \multirow{2}*{$\mathcal{S}$-net} & $93.46\%  $ & $92.31\% $ & $94.69\% $ & $87.84\%$\\
 \rule[-1ex]{0pt}{3.5ex} ~ & ~ & $ \pm 2.79\% $ & $ \pm 4.20\%$ & $\pm 1.31\%$ & $\pm 4.75\%$\\
 \cline{2-6}
 \rule[-1ex]{0pt}{3.5ex}  ~  & \multirow{2}*{$\mathcal{W}$-net} & \bm{$94.76\%$}& \bm{$93.81\% \%$} & $\bm{95.77\% }$ & $\bm{90.07\% }$ \\
 \rule[-1ex]{0pt}{3.5ex}  ~  & ~ & \bm{$ \pm 1.23\% $} & \bm{$ \pm 2.33\%$} & $\bm{ \pm 0.46\%}$ & $\bm{ \pm 2.21\%}$ \\
 \hline
 \rule[-1ex]{0pt}{3.5ex} \multirow{4}*{DiaRetDb1} & \multirow{2}*{$\mathcal{S}$-net} & $95.42\%  $ & $94.76\% $ & $96.10\% $ & $91.25\% $\\
 \rule[-1ex]{0pt}{3.5ex} ~ & ~ & $\pm 0.81\% $ & $ \pm 1.66\%$ & $\pm 0.54\%$ & $ \pm 1.48\%$\\
 \cline{2-6}
 \rule[-1ex]{0pt}{3.5ex} ~  & \multirow{2}*{$\mathcal{W}$-net} & \bm{$95.73\%  $} & \bm{$95.07\%$} & \bm{$96.39\% $} & \bm{$91.81\% $}\\
 \rule[-1ex]{0pt}{3.5ex} ~  & ~ & \bm{$\pm 0.66\% $} & \bm{$\pm 0.82\%$} & \bm{$\pm 0.52\%$} & \bm{$ \pm 1.22\%$}\\
 \hline
 \end{tabular}
 \label{tab:OD segmentation}
 \end{table*}
 
 \begin{table*}[htp!]
 \centering
 \caption{Results of the exudate segmentation using the $\mathcal{S}$-net and $\mathcal{W}$-net on the two datasets}
 \begin{tabular}{c|c|ccc}
 \hline
 \multicolumn{2}{c|}{Experiments} & F1-score & Sensitivity & Precision \\
 \hline
 \rule[-1ex]{0pt}{3.5ex} \multirow{2}*{e\_ophtha\_EX} & $\mathcal{S}$-net & $91.57\% \pm 1.01\% $ & \bm{$92.74\% \pm 1.26\%$} & $90.47\% \pm 1.95\%$ \\
 \cline{2-5}
 \rule[-1ex]{0pt}{3.5ex}          ~       & $\mathcal{W}$-net & \bm{$92.80\% \pm 0.79\% $} & $92.65\% \pm 1.15\%$ & \bm{$92.97\% \pm 1.67\%$} \\
 \hline
 \rule[-1ex]{0pt}{3.5ex} \multirow{2}*{DiaRetDb1} & $\mathcal{S}$-net & $91.25\% \pm 2.59\% $ & $89.57\% \pm 3.31\%$ & $93.06\% \pm 2.86\%$ \\
 \cline{2-5}
 \rule[-1ex]{0pt}{3.5ex}          ~       & $\mathcal{W}$-net & \bm{$94.14\% \pm 1.62\% $} & \bm{$93.26\% \pm 3.27\%$} & \bm{$95.12\% \pm 1.12\%$} \\
 \hline
 \end{tabular}
 \label{tab:exudates segmentation}
 \end{table*}

\begin{figure}[htp!]
 \centering
 \includegraphics[width=.85\linewidth]{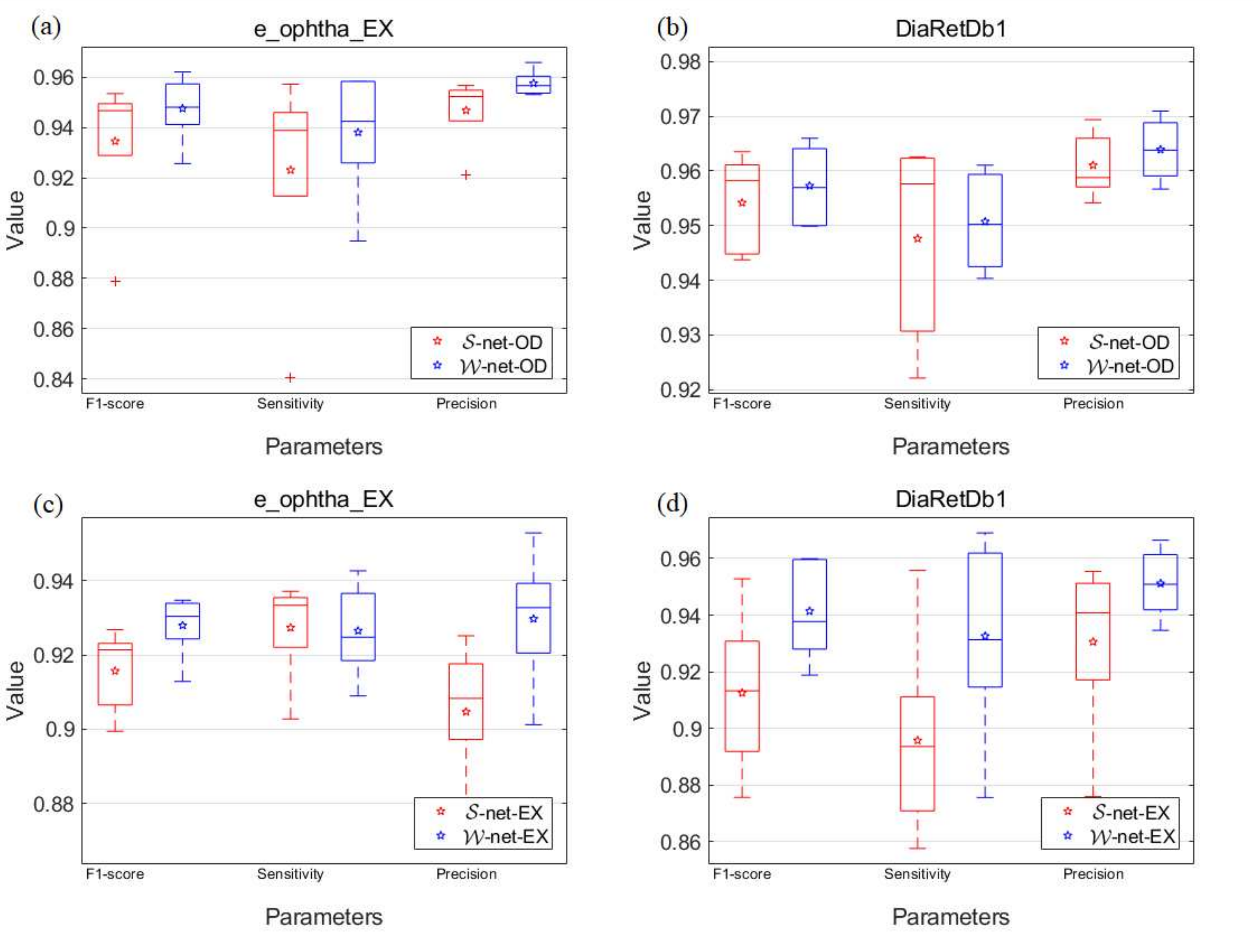}
 \caption{The comparisons of the performance of the $\mathcal{S}$-net and the $\mathcal{W}$-net using five-fold cross-validation. The first row (a and b) showed the comparison of the OD segmentation as tested on the e\_ophtha\_EX and the DiaRetDb1 datasets. The second row (c and d) compared the performance of the exudate segmentation using the same two datasets. }
 \label{fig:box}
 \end{figure}

 \noindent The $\mathcal{W}$-net outperformed the $\mathcal{S}$-net for both OD and exudate segmentation. As shown in Fig.~\ref{fig:box}, on both datasets,  the $\mathcal{W}$-net demonstrated better performance for both OD and exudate segmentation than the $\mathcal{S}$-net, as evaluated using F-1 score, sensitivity, and precision. Particularly, for the exudate segmentation the F-1 score achieved using the $\mathcal{W}$-net was more than 2\% higher than that achieved using the $\mathcal{S}$-net ($p<0.05$) on the two datasets. The F-1 score is a more comprehensive performance metric than the sensitivity or the precision. For the OD segmentation, the $\mathcal{W}$-net also showed better performance than the $\mathcal{S}$-net though the difference was not as significant as the improvement in the exudate segmentation. Table~\ref{tab:OD segmentation} and Table~\ref{tab:exudates segmentation} listed all the values. 
 
\begin{figure}[htp!]
 \centering
 \includegraphics[width=.85\linewidth]{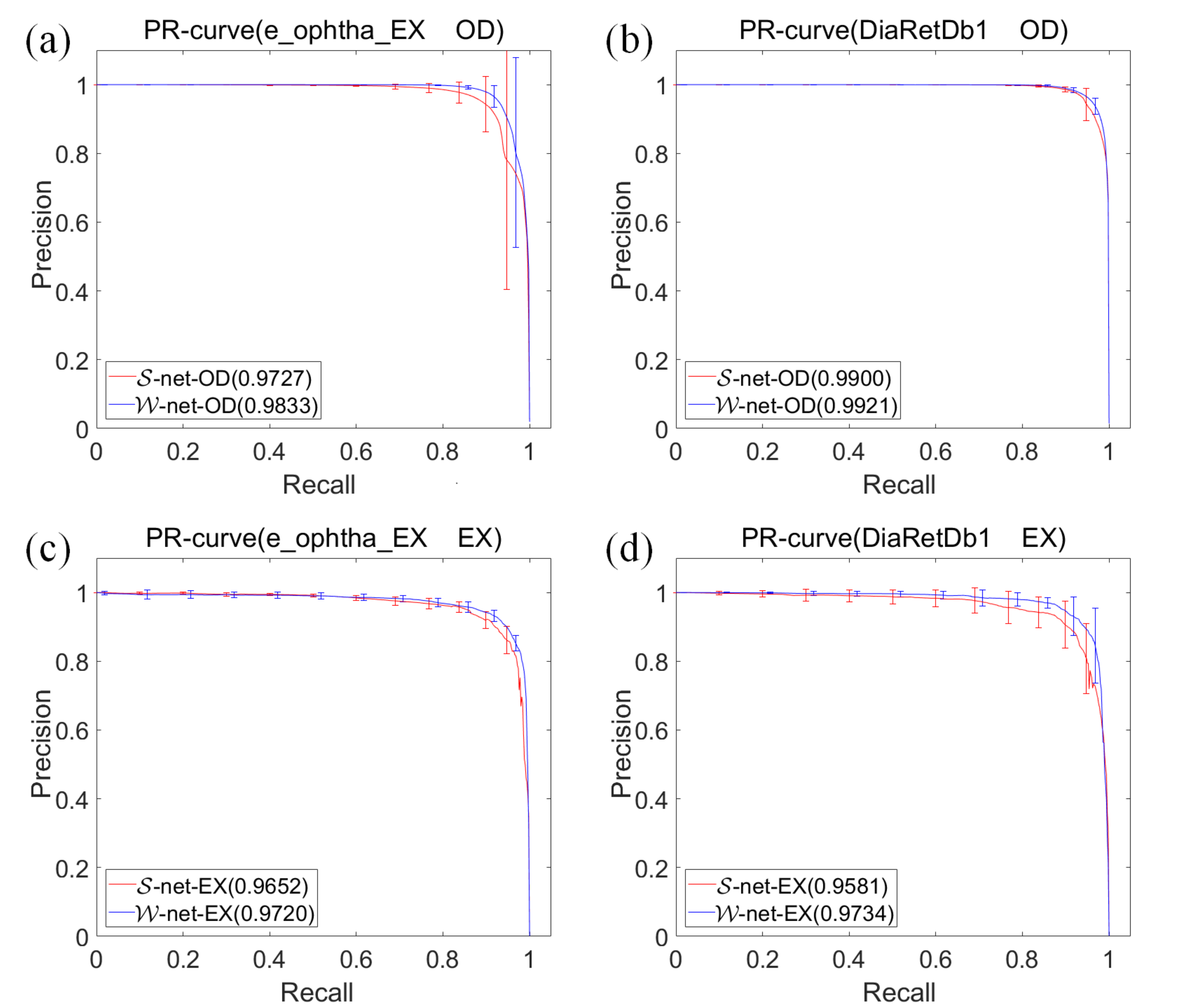}
 \caption{The comparisons of PR curves of the $\mathcal{S}$-net and $\mathcal{W}$-net using five-fold cross-validation. The first row (a and b) showed the comparison of the OD segmentation as tested on the  e\_ophtha\_EX and the DiaRetDb1 datasets. The second row (c and d) compared the performance of the exudate segmentation using the same two datasets. Comparing with $\mathcal{S}$-net, it's obvious that the $\mathcal{W}$-net has achieved higher AUCs of the PR curves in both exudates and optic disc segmentation task in two datasets.}
 \label{fig:pr}
 \end{figure}

\begin{figure}[htpb]
 \centering
 \includegraphics[width=0.8\linewidth]{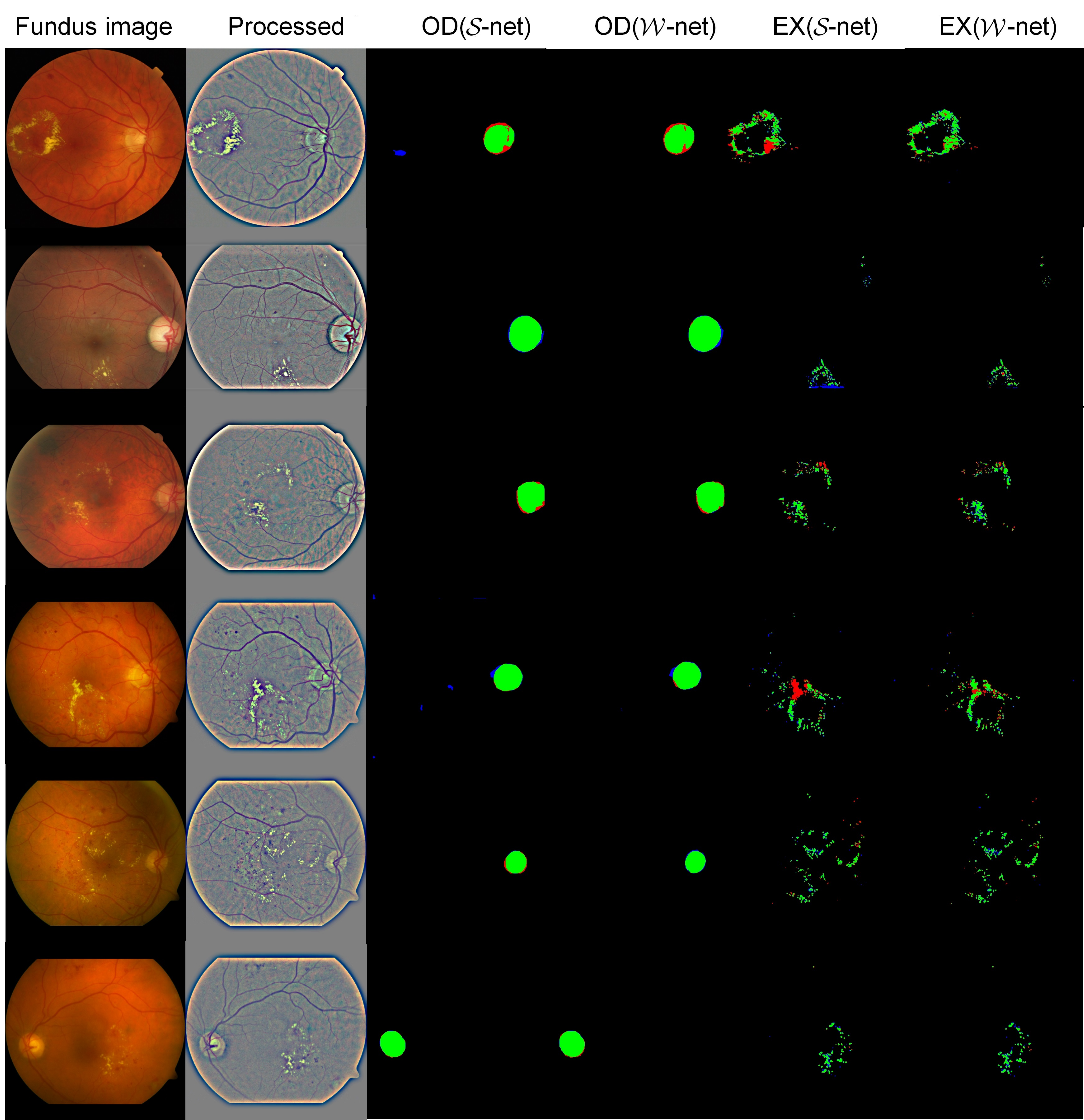}
 \caption{The first three rows displayed images of the e\_ophtha\_EX dataset and the last three rows showed images of the DiaRetDb1 dataset. The first column listed the RGB images. The second column listed the preproccessed images, and the third and forth columns listed the segmented ODs using the $\mathcal{S}$-net and $\mathcal{W}$-net, respectively. The last two columns illustrated the segmentated exudates using the $\mathcal{S}$-net and $\mathcal{W}$-net, respectively. The green areas represented the correctly predicted lesions by algorithms (true positive), the red areas represented lesions not predicted by algorithms (false negative), and the blue areas represented the falsely predicted lesions by the algorithms (false positive).}
 \label{fig:EDdatabase}
 \end{figure}

 Fig.~\ref{fig:pr} showed the precision-recall (PR) curves with the area under curves (AUCs) of the $\mathcal{S}$-net and $\mathcal{W}$-net. Each curve was the average of the five cross-validation results. For OD segmentation, the AUCs using the $\mathcal{S}$-net were 0.9727, 0.9900 on the  e\_ophtha\_EX and the DiaRetDb1 datasets respectively, compared with 0.9833 and 0.9921 using the $\mathcal{W}$-net. For the exudate segmentation, with the $\mathcal{W}$-net the AUCs were 0.9720 and 0.9734 compared with 0.9652 and 0.9581 using the $\mathcal{S}$-net as tested on the e\_ophtha\_EX and the DiaRetDb1 datasets, respectively. The average curve of the $\mathcal{W}$-net completely enclosed the curve of the $\mathcal{S}$-net, demonstrating the better performance of the proposed $\mathcal{W}$-net. Fig.~\ref{fig:EDdatabase} showed the segmentation results of the $\mathcal{S}$-net and $\mathcal{W}$-net on the e\_ophtha\_EX and the DiaRetDb1 datasets.
 
 \begin{table}[htpb]
 \centering
 \caption{Results of the OD segmentation on DRIONS-DB dataset and exudate segmentation on the MESSIDOR dataset using the $\mathcal{S}$-net and $\mathcal{W}$-net}
 \begin{tabular}{c|c|ccc}
 \hline
 \multicolumn{2}{c|}{Experiments} & F1-score & Sensitivity & Precision \\
 \hline
 \rule[-1ex]{0pt}{3.5ex} \multirow{2}*{DRIONS-DB(OD)} & $\mathcal{S}$-net & $89.91\%  $ & $89.27\% $ & $90.36\% $ \\
 \cline{2-5}
 \rule[-1ex]{0pt}{3.5ex}    ~   & $\mathcal{W}$-net & \bm{$95.57\% $} & \bm{$95.50\% $} & \bm{$95.64\%$} \\
 \hline
 \rule[-1ex]{0pt}{3.5ex} \multirow{2}*{MESSIDOR(EX)} & $\mathcal{S}$-net & $91.36\% $ & $92.94\% $ & $89.83\% $ \\
 \cline{2-5}
 \rule[-1ex]{0pt}{3.5ex}          ~       & $\mathcal{W}$-net & \bm{$92.66\% $} & \bm{$95.32\% $} & \bm{$90.15\%$} \\
 \hline
 \end{tabular}
 \label{tab:robust}
 \end{table}
 From the images, we can see that the false positive and false negative of the $\mathcal{S}$-net is more than the $\mathcal{W}$-net. The green areas, namely true positive, of the $\mathcal{W}$-net takes up a greater proportion than that of the $\mathcal{S}$-net. From the figure, it is intuitive to see that the $\mathcal{W}$-net achieved better performance than the $\mathcal{S}$-net.
 
 \subsection{The $\mathcal{W}$-net improved the the robustness and generalization of the network}
 \noindent To compare the robustness and generalization properties of the $\mathcal{S}$-net and $\mathcal{W}$-net, we applied the trained $\mathcal{S}$-net and $\mathcal{W}$-net to test OD and exudates segmentation on the DRIONS-DB and MESSIDOR datasets, respectively. Both the DRIONS-DB and MESSIDOR datasets were not applied to train the models. Table~\ref{tab:robust} showed the performance of the $\mathcal{S}$-net and $\mathcal{W}$-net on the two datasets. From the table, it is clear that the $\mathcal{W}$-net is superior to the $\mathcal{S}$-net for the performance of the OD and exudates segmentation. The $\mathcal{W}$-net surpassed the $\mathcal{S}$-net more than 5.6\% of F1-score on OD segmentation on the DRIONS-DB dataset and exceeded more than 1.3\% on exudates segmentation on the MESSIDOR dataset. It is conclusive that the robustness and generalization of the $\mathcal{W}$-net are much better than the $\mathcal{S}$-net.

 \section{Discussion}
 \noindent Simultaneous segmentation of multiple anatomical structures is of great importance in medical image analysis. Most existing DCNNs apply independent networks for each individual task and completely ignore the correlation information among tasks. 
 To utilize the correlation among different tasks, we proposed a $\mathcal{W}$-net to simultaneously segment both OD and exudates. These two tasks shared the same encoder path, which greatly reduced the risk of overfitting~\cite{ruder2017overview}. Indeed, the work by Baxter \textit{et al.}~\cite{baxter1997bayesian} showed that the risk of overfitting the shared parameters was order $N$ smaller than that of fitting the task-specific parameters, where $N$ was the number of tasks.
 
 Another advantage of the MTL based $\mathcal{W}$-net was the significant reduction of the fitting parameters. 
 The $\mathcal{W}$-net had 2.9M trainable parameters, while the two $\mathcal{S}$-nets had 4.1M parameters. 
 The $\mathcal{W}$-net was not only more efficient in the training stage, but also faster in the testing stage than the $\mathcal{S}$-net. For instance, to finish an image using the $\mathcal{W}$-net need 0.15 seconds while it took 0.22 seconds using two $\mathcal{S}$-net sequentially on an Intel Core i7-7700 CPU with Nvidia GeForce GTX 1080 Ti. This is important where long computation time is prohibitive.  
 \begin{table*}[htbp]
 \centering
 \small
 \setlength\tabcolsep{9pt}
 \caption{Comparison of the performances of the OD segmentation of our method with previous published methods on two public datasets}
 \begin{tabular}{llcccc}
 \hline
 \rule[-1ex]{0pt}{3.5ex} Dataset & Methodology & $\eta$ & F1-score &Sensitivity& Precision \\
 \hline
 \rule[-1ex]{0pt}{3.5ex} DiaRetDb1 & Walter {\textit{et al.}}\cite{walter2002contribution} & $36.97\%$ & - &65.69\% &-  \\
 \rule[-1ex]{0pt}{3.5ex}  & Abdullah {\textit{et al.}}\cite{abdullah2016localization} & $85.1\%$ & 89.10\% &85.08\% &-  \\
 \rule[-1ex]{0pt}{3.5ex}  & Morales {\textit{et al.}}\cite{morales2013automatic} & - & $89.30\%$ &-& -  \\
 \rule[-1ex]{0pt}{3.5ex}  &Proposed & $\bm{91.81\%}$ & $\bm{95.73\%}$ &$\bm{95.07\%}$& $\bm{96.39\%}$ \\
 \hline
 \rule[-1ex]{0pt}{3.5ex} DRIONS-DB &  Walter {\textit{et al.}}\cite{walter2002contribution} & - & 68.13\% & - &- \\
 \rule[-1ex]{0pt}{3.5ex} & Zahoor {\textit{et al.}}\cite{zahoor2017fast} & $88.6\%$ & - &93.84\% &- \\
 \rule[-1ex]{0pt}{3.5ex} & Abdullah {et  al.}\cite{abdullah2016localization} & $85.1\%$ & 91.02\% &85.1\% &- \\
 \rule[-1ex]{0pt}{3.5ex}  &Proposed & $\bm{91.52\%}$& $\bm{95.57\%}$ &$\bm{95.50\%}$&$ \bm{95.64\%}$ \\
 \hline
 \end{tabular}
 \label{tab:comparison_others_OD}
 \end{table*}
 
 \begin{table*}[htbp]
 \centering
 \setlength\tabcolsep{7pt}
 \caption{Comparison of the performance of the exudate segmentation of the $\mathcal{W}$-net with previous published methods on the two public datasets}
 \begin{tabular}{llcccc}
 \hline
 \rule[-1ex]{0pt}{3.5ex} Database & Methodology &  Accuracy & Specificity & Sensitivity & Precision  \\
 \hline
 \rule[-1ex]{0pt}{3.5ex} e\_ophtha\_EX & Moazam {\textit{et al.}}\cite{fraz2017multiscale} & $89.25\%$ & $94.60\%$ & \rule[-1ex]{0pt}{3.5ex} $81.20\%$ & $90.91\%$ \\
 \rule[-1ex]{0pt}{3.5ex}  & Zhang {\textit{et al.}}\cite{zhang2014exudate} & $--$ & $--$ & $74\%$ & $79\%$  \\
 \rule[-1ex]{0pt}{3.5ex}  & Proposed & $\bm{99.97\%}$ & $\bm{99.99\%}$ & $\bm{92.65\%}$ & $\bm{92.97\%}$ \\
 \hline
 \rule[-1ex]{0pt}{3.5ex} DiaRetDb1 & B Harangi {\textit{et al.}}\cite{harangi2014automatic} & - & - & $85\%$ & $84\%$ \\
 \rule[-1ex]{0pt}{3.5ex} & Moazam {\textit{et al.}}\cite{fraz2017multiscale} & $87.72\%$ & $81.25\%$ & $92.42\%$ & $87.14\%$ \\
 \rule[-1ex]{0pt}{3.5ex} & Zheng {\textit{et al.}}\cite{zheng2018detection} & $99.97\%$ & $99.98\%$ & \bm{$93.94\%$} & $91.02\%$  \\
 \rule[-1ex]{0pt}{3.5ex} & Proposed & $\bm{99.98\%}$ & $\bm{99.99\%}$ & $93.26\%$ & $\bm{95.12\%}$  \\
 \hline
 \end{tabular}
 \label{tab:comparison_others_EX}
 \end{table*}
 
 \begin{figure}[htpb]
 \centering
 \includegraphics[width=0.88\linewidth]{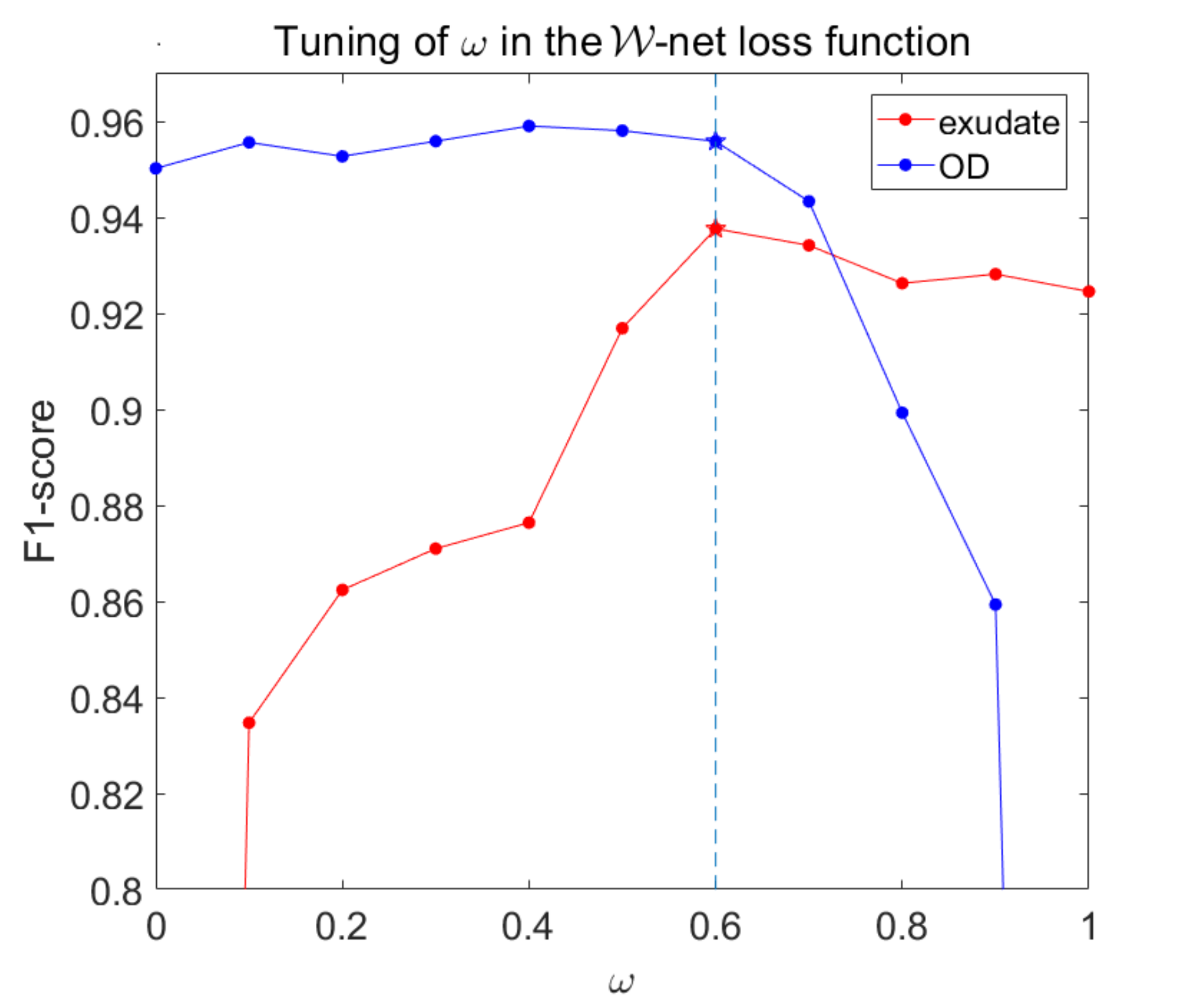}
 \caption{The F1-score of exudate and OD segmentation vs different loss weight on the DiaRetDb1 dataset.}
 \label{fig:searchw}
 \end{figure}
 
 We compared our results with recently published literature on OD (Table~\ref{tab:comparison_others_OD}) and exudate segmentation (Table~\ref{tab:comparison_others_EX}). Since most of the published work on OD segmentation used DRIONS-DB dataset to evaluate the performance of the segmentation algorithms, we applied the $\mathcal{W}$-net on the same dataset for comparison. As shown in Table~\ref{tab:comparison_others_OD}, our proposed $\mathcal{W}$-net outperformed other methods on both datasets by a large margin as evaluated by either $\eta$, F1-score, sensitivity or precision.
 
 Table~\ref{tab:comparison_others_EX} showed the comparison on the exudate segmentation. We made sure that all the studies listed used the same lesion level evaluation method~\cite{wolf2006object} as used in our study to keep a fair comparison. From the table, it was clear that our proposed $\mathcal{W}$-net surpassed most of the methods published except the method proposed by Zheng \textit{et al.}~\cite{zheng2018detection} in sensitivity. However, we noticed that Zheng \textit{et al.} applied an ensemble method instead of using a single network. 
 
 As illustrated in the reference~\cite{kendall2017multi}, the performance of a multi-task deep convolutional neural network is strongly dependent on the relative weight between the loss of each task. This is manifested in the tuning of the $\omega$ parameter in Eq.~\ref{Eq:total_loss1}. Kendall \textit{et al.}~\cite{kendall2017multi} proposed a method to choose the optimal weights by considering the homoscedastic uncertainty of each task. In this study, we applied a more conventional method by scanning the value of $\omega$ from 0 to 1 with an incremental step size of 0.1. As shown in Fig.~\ref{fig:searchw}, the $\mathcal{W}$-net achieved best performance of both OD and exudate segmentation on DiaRetDb1 dataset when $\omega$ was set to 0.6.
 
 \section{Conclusion}
 \noindent In summary, we sought to develop a deep neural network combined with MTL for simultaneous exudate and OD segmentation. Based on the $\mathcal{S}$-net, an improved version of U-net, we developed a $\mathcal{W}$-net, which demonstrated excellent performance when compared with previous studies. Compared with applying two $\mathcal{S}$-net sequentially, $\mathcal{W}$-net was more efficient and robust. The recent study by Kendall \textit{et al.}~\cite{kendall2017multi,li2016deepsaliency,li2014heterogeneous} came to the same conclusion that by optimally weighting loss terms in the multi-task loss function, the multi-task model outperformed separate models trained individually on each task. This method can be readily applied to more than two tasks. One of the difficulties of applying such method was due to the lack of multi-labeled datasets, which is a general problem in the application of deep neural networks. Though in this study, both tasks were applied on the fundus images, MTL method could be used to process images from different modalities, for example, images from optical coherence tomography (OCT) in combination with ultrasound. 
 
 \section{Acknowledgment}
 The work is partially supported by the National Natural Science Foundation of China (Key Program) under grand No.U19B2044. It is supported by the GPU Computing Cluster of the data center, School of Information Science and Technology, University of Science and Technology of China. 

\bibliographystyle{IEEEtran}
\bibliography{report}

\EOD
\end{document}